\documentclass[showpacs,pre,twocolumn,amsmath,amssymb,epsfig]{revtex4-1}

\usepackage{color}

\usepackage{graphicx}
\usepackage{epsfig}
\usepackage{dcolumn}
\usepackage{bm}
\newcommand     {\beq}[1]         { \begin{equation} #1 \end{equation} }
\newcommand     {\beqa}[1]        { \begin{eqnarray} #1 \end{eqnarray} }

\begin{document}

\title{Approach to failure in porous granular materials under compression} 

 \author{Ferenc Kun${}^{1}$\email{ferenc.kun@science.unideb.hu}, Imre
Varga${}^{2}$, Sabine Lennartz-Sassinek${}^{3,4}$, Ian G.\ Main${}^{4}$} 
\affiliation{$^1$Department of Theoretical Physics, University of Debrecen,
P.O. Box 5, H-4010 Debrecen, Hungary}
\affiliation{$^2$Department of Informatics Systems and Networks, University of Debrecen,
P.O. Box 12, H-4010 Debrecen, Hungary}
\affiliation{$^3$School of Geosciences, University of Edinburgh,  EH9 3JL
Edinburgh, UK}
\affiliation{$^4$Institute for Geophysics and Meteorology, 
University of Cologne, Cologne, Germany}

\begin{abstract}
We investigate the approach to catastrophic failure in a 
model porous granular material undergoing uniaxial compression. A discrete
element computational model is used to simulate both the micro-structure
of the material and the complex dynamics and feedbacks involved in local
fracturing and the production of crackling noise.  Under
strain-controlled loading micro-cracks initially nucleate in an
uncorrelated way all over the sample. As loading proceeds the damage
localizes into a narrow damage band inclined at 30-45 degrees to the
load direction. Inside the damage band the material is crushed into a
poorly-sorted mixture of mainly fine powder hosting some larger
fragments. The mass probability density distribution of particles in the
damage zone is a power law of exponent 2.1, similar to a value of 1.87 inferred
from observations of the length distribution of wear products (gouge) in natural
and laboratory faults. Dynamic bursts of radiated energy,
analogous to acoustic emissions observed in laboratory experiments on porous sedimentary rocks, 
are identified as correlated trails or cascades of local ruptures that
emerge from the stress redistribution process.  As the system approaches
macroscopic failure consecutive bursts become progressively more
correlated. 
Their size distribution is also a power law, with an equivalent
Gutenberg-Richter ‘b-value’ of 1.22 averaged over the whole test, ranging from 3
down to 0.5 at the time of failure, all similar to those observed in laboratory
tests on granular sandstone samples.  The formation of the damage band itself is
marked by a decrease in the average distance between consecutive bursts and an
emergent power law correlation integral of event locations with a correlation
dimension of 2.55, also similar to those observed in the laboratory (between
2.75 and 2.25).
\end{abstract}
\maketitle

 \section{Introduction}

The compressive failure of porous sedimentary rocks is important in a range of
applications in geophysics and engineering. They are used as building materials,
and their failure mechanisms control the nature and timing of natural or induced
hazards such as landslides and earthquakes in such materials 
\cite{sammonds_role_1992,ojala_2004,Heap201171,main_fault_2000,ROG:ROG1468}.
In these and other cohesive granular materials failure occurs by the
intermittent
nucleation, propagation and coalescence of micro-cracks that generate acoustic
emissions, one of the main sources of information about the microscopic dynamics
of such failure processes
\cite{lockner_nature_1991,lockner_1993,sammonds_role_1992}.
Laboratory experiments have revealed that both the spatial structure of
damage and the statistical and dynamical features of the time series 
of the corresponding acoustic events undergo a complex 
evolution when approaching macroscopic failure 
\cite{sammonds_role_1992,ROG:ROG1468,schorlemmer_variations_2005}. 
Quantitative understanding 
of this evolution process is indispensible to design forecasting strategies 
for imminent catastrophic failure \cite{ROG:ROG1468}.
Triaxial loading experiments on Earth materials, well representing field
conditions, 
showed that the beginning of the loading process is dominated by random
nucleation of micro-cracks.
However, in the vicinity of failure correlations dominate, i.e.\ damage was
found 
to localize in narrow bands which gradually broaden 
\cite{sammonds_role_1992,ojala_2004,Heap201171,main_fault_2000,ROG:ROG1468}. 
The integrated statistics of the energy
of acoustic emissions, accumulating all the events up to failure, is
characterized by a power-law distribution $p(E)\sim E^{-B-1}$ 
where the exponent $B$ shows some degree of robustness with respect to
material properties \cite{main_fault_2000,ROG:ROG1468}. 
However, in narrow time/strain
windows a systematic decrease of the exponent $B$ was observed 
as failure is approached \cite{sammonds_role_1992}.
These temporal changes have been suggested as possible diagnostic signatures
of imminent catastrophic failure \cite{ROG:ROG1468,amitrano_2012_epjst}.

The statistics of breaking bursts are usually investigated in the framework of 
stochastic lattice models, which are based on regular lattices of springs
\cite{alava_statistical_2006}, 
beams \cite{timar_crackling_2011}, fibers \cite{pradhan_failure_2010}, or 
fuses \cite{alava_statistical_2006,alava_role_2008}. They have the advantage of
allowing
for a straightforward identification of breaking avalanches. However, they imply
strong simplifications of the micro-structure of materials and on the dynamics
of local breakings. Furthermore, they are typically capable of handling only tensile
loading conditions. Stochastic lattice models have all reproduced the integrated power
law statistics 
of burst size and revealed interesting effects of the amount of disorder 
\cite{alava_statistical_2006,alava_role_2008,hidalgo_universality_2008,
PhysRevE.87.042816,hidalgo_avalanche_2009-1}, friction
\cite{girard_failure_2010,amitrano_2012_epjst}, and range of 
load redistribution on the value of the exponent
\cite{PhysRevE.87.042816,raischel_local_2006}.
As an alternative, the discrete element modeling (DEM) approach provides a more
realistic treatment especially for cohesive granular materials which are
inherently discrete
\cite{cundall_discrete_1979,kun_study_1996,daddetta_application_2002,
potyondy_bonded-particle_2004,hentz_discrete_2004,ergenzinger_discrete_2010}.
In the framework of DEM both the micro-structure of the material and the
dynamics of fracture can be better captured. 
Hence, it has successfully reproduced the spatial structure of damage
under several loading conditions
\cite{cundall_discrete_1979,kun_study_1996,daddetta_application_2002,
potyondy_bonded-particle_2004,hentz_discrete_2004,ergenzinger_discrete_2010,
carmona_fragmentation_2008,timar_new_2010}. 

Recently, we have introduced a method in DEMs
to investigate crackling noise generated by cascades of micro-fractures similar
to sources 
of acoustic emission in experiments \cite{ourpaperinprl}. 
Crackling bursts are identified as correlated 
trails of breaking particle contacts which made it possible to decompose the
process of 
damage accumulation into a time series of bursts. By simulating uniaxial
compression 
of cylindrical samples of sedimentary rocks we have shown that our DEM
reproduces 
all qualitative features of the integrated statistics of the time series of
acoustic events
\cite{ourpaperinprl}.
Here we apply our model to investigate how the system approaches macroscopic
failure 
by analyzing both the spatial structure of damage and the variation of the power
law 
statistics of burst sizes in the approach to failure. A connection of the two is
established by 
investigating the spatial correlation of consecutive bursts.

 \section{Discrete element model of geomaterials}
We briefly summarize the main steps of the construction of our
discrete element model (DEM) which captures the essential ingredients 
of both the heterogeneous micro-structure of sedimentary rocks and the 
dynamics of the breaking process. The model has been used in Ref.\
\cite{ourpaperinprl}
to investigate the integrated statistics of crackling avalanches, 
i.e.\ the average properties in periods both up to and beyond 
the peak stress, respectively. Here we will examine the temporal 
evolution of the microstructural and mechanical properties 
in the approach to failure in more detail.

 \subsection{Heterogeneous micro-structure}
\begin{figure}
 \begin{center}
 \epsfig{ bbllx=0,bblly=0,bburx=350,bbury=415,file=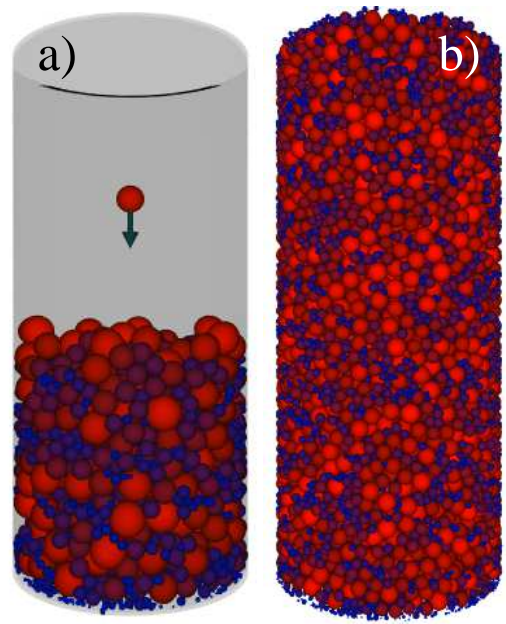,
 width=7.0cm}
 \end{center}
 \caption{
 (Color online) $(a)$ Preparation of the sample by sedimenting spherical particles
 with radius $R$ sampled from a probability distribution $p(R)$. 
 The color code corresponds to the radius of the particles such that the
 smallest particles are dark blue while the biggest ones have light red color.
 $(b)$ A complete cylindrical sample of $20000$ particles used in the simulations
 of the fracture process. 
 }
 \label{fig:sediment}
 \end{figure}

 In order to represent the structure of sedimentary rocks in a discrete element
framework
 we sedimented spherical shaped particles in a cylindrical container of diameter $D_0$
 and height $H_0$.
 During the sedimentation process particles fall one-by-one at random lateral position
 on the top of the growing particle layer and dissipate their
 kinetic energy by colliding with other particles and also with the 
 wall of the container. The radius of particles $R$ is randomly generated
 according to a probability density $p(R)$.
 Figure \ref{fig:sediment}$(a)$ illustrates 
 the procedure of sample generation where the color code represents the 
 particle size. The origin of the coordinate system is in the middle of the bottom plate
 of the cylinder and the $z$ axis is aligned with the symmetry axis.

Molecular dynamics simulations were carried in order to generate the 
ballistic motion of particles of mass $m_i$  ($i=1,\ldots , N$) under the action of the gravitational force
$\vec{F}_i^g = m_i\vec{g}$, where $\vec{g}$ is the gravitational acceleration.
The particles fall with zero initial speed and find their equilibrium position in the random packing through a 
sequence of collisions. In the simulation 
we apply a soft particle contact model, i.e.\ particles overlap
when they are pressed against each other which then gives rise to a repulsive
force between them. Two particles $i$ and $j$ with radii $R_i$, $R_j$ 
and position vectors $\vec{r}_i$, $\vec{r}_j$ are in contact during their
motion when the overlap
distance $\xi = R_i+R_j-r_{ij}$ is positive.
Here $r_{ij}$ denotes the distance of the particles $r_{ij}=|\vec{r}_{ij}|$ 
with $\vec{r}_{ij}=\vec{r}_i-\vec{r}_j$ pointing from particle $j$ to $i$.
The interaction of the particles is described by the Hertz contact law
including also viscoelastic dissipation \cite{poschel_grandyn_2005}:
the contact force $\vec{F}^c_{ij}$ exerted by particle $j$ on $i$ 
is expressed in terms of the overlap distance $\xi$ as
\beq{
\vec{F}^c_{ij} = -k^p_{ij}(\xi^{3/2}+A\sqrt{\xi}\dot{\xi})\vec{n}_{ij}.
\label{eq:hertz}
} 
The contact stiffness $k^p_{ij}$ depends 
on the material and geometrical properties of particles as $k^p_{ij}=2E_p\sqrt{R_{eff}}/3(1-\nu_p^2)$,
where $1/R_{eff}=1/R_i + 1/R_j$, furthermore, $E_p$ and $\nu_p$ denote the Young modulus
and Poisson ratio of particles' material. The unit vector $\vec{n}_{ij}=\vec{r}_{ij}/|\vec{r}_{ij}|$ 
characterizes the orientation of the contact. 
In the force law Eq.\ (\ref{eq:hertz}) dissipation of
the kinetic energy is ensured by the rate $\dot{\xi}$ dependent term, where 
the parameter $A$ captures the viscoelastic properties of the material.
For simplicity, during the sedimentation process no tangential component
of the contact force was considered. In order to generate samples 
with the required overall geometrical shape, bouncing particles interact 
with the container wall, as well. The force exerted by the wall on particle $i$ 
has the form 
\beq{
\vec{F}^w_{i} = -k^w_i(R_i+|\vec{r}_i^r|-D_0/2)\vec{r}_i^r/|\vec{r}_i^r|, 
}
where $\vec{r}_i^r$
denotes the radial component of the position vector $\vec{r}_i$ so that
$\vec{F}^w_{i}$ points towards the symmetry axis of the cylinder.
The time evolution of sedimentation was generated by numerically solving 
the equation of motion of single particles
\beqa{
m_i\ddot{\vec{r}}_{i} &=& \sum_j \vec{F}_{ij}^c + \vec{F}^w_{i} + \vec{F}^g_{i},
\label{eq:eom}
}
where the sum over $j$ in the first term of the right hand side runs over all
contacts of particle $i$. Eq.\ (\ref{eq:eom}) is solved for each particle 
independently assuming all other particle positions to be frozen.
In the simulations a 3$rd$ order Predictor-Corrector
scheme was used for the numerical solution of Eq.\ (\ref{eq:eom}) which
ensures stability and high precision \cite{allen_computer_1984}. 

The single-particle sedimentation technique has several advantages for the preparation of 
random particle packings: the bouncing particle
can solely interact with the top layer in the container, hence, 
those particles which are located inside the sediment can be considered fixed, 
and hence, they are not considered in contact searching.
Since these particle positions are not updated the simulation time linearly scales
with the particle number.
After the energy of the sedimenting particle droped below a small threshold value the motion of the particle 
was stopped. In such a configuration the particle has typically small overlaps
with the surrounding ones due to the action of gravity. In order to ensure a stress free initial 
packing we slightly displaced the sedimented particle along the direction of the sum of 
contact forces until all overlaps were removed without changing the radius of the particle.  
The efficiency of the sedimentation techniques made it possible to generate packings
of $10^6$ particles with about 6 hours CPU time on a single core of an Intel Xeon (6 cores)
processor.
The disadvantage of the technique is that the size distribution of particles
$p(R)$ cannot be arbitrarily broad. Very large particles may prevent sedimenting
particles to fill holes around them which may create an unphysical void structure. 
In the opposite limit, very small particles have a high chance to bounce inside
the holes between the larger ones sedimenting to the bottom of the container.
This way small particles would accumulate at the bottom, while the very large ones 
would stay at the top of the packing creating an unphysical micro-structure.

The radius of particles $R$ was sampled from a log-normal distribution
 \beq{
 p(R) \sim \exp{\left[-\frac{(\ln R - \ln \overline{R})^2}{2\sigma_R^2}\right]},
 \label{eq:lognorm}
 }
which gives a reasonable description of 
 the statistics of particle sizes for various types of earth materials in 
 the range of large particles (see e.g.\ the particle mass distribution prior to
faulting
 in Figure 7 of Ref.\ \cite{Mair200025}). 
To avoid size segregation of particles during sedimentation we set the range of
radius 
 $R_{min}\leq R \leq R_{max}$ such that the ratio $R_{max}/R_{min}=20$ is fixed.
 Then $\overline{R}$ is chosen as $\overline{R}=5R_{min}$ to have the maximum of
$p(R)$
 nearly in the middle of the $[\log R_{min}, \log R_{max}]$ interval. 
 In the model the smallest particle radius $R_{min}$ is used as characteristic 
 length scale in terms of which all other lengths are expressed.
 The diameter $D_0$ and height $H_0$ of the cylinder were chosen to be
 $D_0=438.57R_{min}$ and $H_0=1008.71R_{min}$, which yields an aspect 
 ratio $H_0/D_0\approx 2.3$ as in the experiments of Ref.\ \cite{Mair200025}.
 With this geometrical setup the number of particles used in the simulations 
 fluctuates in a narrow interval around 20000.
 The final cylindrical sample with random heterogeneous micro-structure is illustrated
 in Fig.\ \ref{fig:sediment}$(b)$. 
 
\begin{figure}
 \begin{center}
 \epsfig{ bbllx=0,bblly=0,bburx=720,bbury=610,file=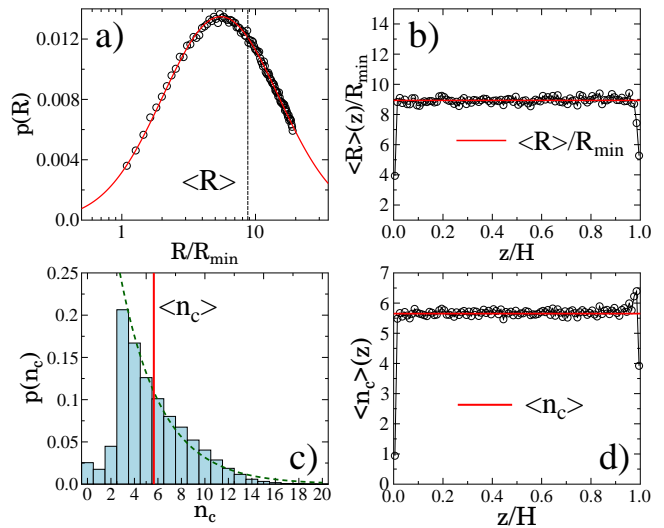,
 width=8.5cm}
 \end{center}
 \caption{
 (Color online) Analysis of the structure of the sediment. $(a)$ 
 The probability distribution of the particle radius in the cylindrical sample.
 The numerically measured distribution (symbols) has a very good agreement with
the 
 analytic curve (continuous line). 
 $(b)$ The average radius $\left<R\right>$ of particles as a function of height
$z$
 inside the cylinder. The value of $\left<R\right>$ is nearly constant and it is
equal 
 to the sample average.
 $(c)$ Probability distribution of the contact number $n_c$ of particles. 
 For $n_c\geq 3$ the distribution can be well described by an exponential
 form (green dashed line). $(d)$ The average contact number $\left<n_c\right>$ proved 
 to be independent of the height along the cylinder axis.
 }
 \label{fig:packing}
 \end{figure}

 Figure \ref{fig:packing} presents the results of the analysis of the
 packing structure of the sample. In Fig.\ \ref{fig:packing}$(a)$
 the size distribution of particles obtained numerically follows perfectly 
 the analytic log-normal form. In order to test the homogeneity of the packing 
 we determined the average particle radius $\left<R\right>$ as a function of
the 
 $z$ coordinate measured from the bottom of the container.  
 Fig.\ \ref{fig:packing}$(b)$ shows that the average size is constant 
 $\left<R\right>/R_{min}\approx 8.9$ along the cylinder axis and it is equal 
 to the value obtained by averaging over the complete sample. Another important
 characteristic of the micro-structure is the number of contacts $n_c$ 
 of the particles. Fig.\ \ref{fig:packing}$(c)$ demonstrates that the histogram 
 of contact numbers $p(n_c)$ can be well described by an exponential form 
 for $n_c\geq 3$. A small fraction of zero contacts occurs 
 due to some very fine particles $R\approx R_{min}$ which sediment to the
 bottom of the container and do not touch any other particles.
 Contact numbers $n_c=1,2,3$ typically occur along the surface of the sample
while
 bulk particles are characterized by higher values of $n_c$. 
 The exponential form of $p(n_c)$ and the value of the average contact 
 number $\left<n_c\right> \approx 5.8$ are in a reasonable agreement with
measurements on Earth materials \cite{Mair200025,turcotte_fractals_1997}. 
 Fig.\ \ref{fig:packing}$(d)$ shows that the average
 contact number $\left<n_c\right>$ does not depend on the position $z$ along the
axis of the cylinder and is equal to the sample average of contact numbers. 
The good quality of the tests in Fig.\ \ref{fig:packing}$(b)$ and $(d)$ implies a 
high degree of homogeneity of the sample with a uniformly-random heterogeneous micro-structure.
 
\subsection{Cohesive granular material with breakable contacts}
 In order to capture the cohesive interaction between the particles,
 first we carry out a Delaunay tetrahedrization with the position of particles 
 in three-dimensions. Cementation between grains is represented 
 such that the particles are coupled by beam elements along the edges 
 of tetrahedrons. Physical properties of the beams are determined by the underlying random 
micro-structure of the particle packing: the equilibrium length $l_{ij}^0$
of the beam between particles $i$ and $j$ is the distance of the particle centers 
in the initial configuration $l_{ij}^0 = |\vec{r}_i^0-\vec{r}_j^0|$, while
the beam cross-section $S_{ij}$ is determined by the particle radii as 
$1/S_{ij} = 1/(R_{i}^2\pi) + 1/(R_{j}^2\pi)$. 
It follows that the heterogeneous 
micro-structure of the particle packing gives rise to randomness of the beam geometry
which then shows up in the values of the physical quantities, e.g.\ stiffness and
moduli of beams, as well.

The beam dynamics we implemented is based on Euler-Bernoulli beams as described in Refs.\
\cite{poschel_grandyn_2005,carmona_fragmentation_2008,PhysRevE.86.016113} 
 as the three-dimensional generalization of the approach of
 Refs.\ \cite{kun_study_1996,daddetta_application_2002}. 
For the quantitative characterization of the deformation of beams a local coordinate
system is attached to each particle at the beam ends. 
As the particles undergo translational and rotational motion the
beams suffer elongation, compression, shear, and torsion resulting in forces and torques. 
The axial force $\vec{F}^b_{ij}$ exerted by the beam connecting 
particles $i$ and $j$ on particle $i$ is expressed in terms of the beam elongation 
$\Delta l_{ij}=r_{ij}-l_{ij}^0$ in the form
\beq{
\vec{F}^b_{ij}=-k^b_{ij}\Delta l_{ij}\vec{n}_{ij}.
\label{eq:bforce}
}
The beam stiffness $k^b_{ij}$ depends on the Young modulus $E_b$ and on the geometrical
features of the beam as $k^b_{ij}=E_bS_{ij}/l_{ij}^0$. 
A dissipative component of the force is also added to Eq.\ (\ref{eq:bforce}) similar to the one
used in the packing generation. The flexural forces and torques can be determined by
tracing the change of the orientation of beam ends with respect to the body fixed coordinate
system $\vec{e}_x^b,\vec{e}_y^b,\vec{e}_z^b$ of the particles, where  $\vec{e}_x^b$ 
is aligned with the beam orientation
\cite{poschel_grandyn_2005}. In a simple case when both beam ends 
rotated around the $\vec{e}^b_z$
axis of the body fixed system by angles $\Theta_i^z$  and $\Theta_j^z$ the resulting force 
and torque acting on particle $i$ can be cast into the form \cite{poschel_grandyn_2005}
\beqa{
\vec{Q}^{z,b}_i &=& 3E_bI_{ij}\frac{\Theta_i^z+\Theta_j^z}{(l_{ij}^0)^2}\vec{e}_y^b,  \\
\vec{M}^{z,b}_i &=& E_bI_{ij}\frac{\Theta_i^z+\Theta_j^z}{l_{ij}^0}\vec{e}_z^b + 
\left(\vec{Q}^{z,b}_i\times |\vec{r}_{ij}|\vec{e}_x^b \right), 
}
where $I_{ij}$ denotes the beam's moment of inertia. Torsion arises due to the relative 
rotation around the $\vec{e}_x^b$ axis which gives rise to the moment
\beq{
\vec{M}^{x,b}_i = G_{ij}I^t_{ij} \frac{\Theta_i^x-\Theta_j^x}{l_{ij}^0}\vec{e}_x^b.
}
Here $G_{ij}$ is the shear modulus of the beam, while $I^t_{ij}$ denotes 
the torsional moment of inertia calculated with respect to the beam axis.
Beam forces and torques must be transformed to the global coordinate system of the particle
packing where the equation of motion is solved numerically for the translational and rotation 
degrees of freedom \cite{allen_computer_1984}. The same 3rd order Predictor-Corrector
solver is used for the simulations as for the generation of the initial particle packing 
including quoternions for the representation of rotations \cite{allen_computer_1984}.

In order to capture crack formation in the model, beams break when  they get
overstressed
during the evolution of the system. The breaking of a beam is mainly caused 
by stretching and bending, hence, we implemented a von-Mises type breaking
criterion widely used in the literature 
\cite{kun_study_1996,daddetta_application_2002,carmona_fragmentation_2008}
 \beq{
\left(\frac{\varepsilon_{ij}}{\varepsilon_{th}}\right)^2 + 
\frac{max(|\Theta_i|, |\Theta_j|)}{\Theta_{th}} > 1.
\label{eq:breaking}
}
Here the axial strain $\varepsilon_{ij}$ of the beam is determined 
as $\varepsilon_{ij}=\Delta l_{ij}/l_{ij}^0$, while $\Theta_i$ and
$\Theta_j$
are the generalized bending angles of the two beam ends.
The first and second terms of Eq.\ (\ref{eq:breaking}) represent
 the contributions of stretching and bending deformations, respectively. The value
 of the breaking thresholds $\varepsilon_{th}$ and $\Theta_{th}$ control the
 relative importance of the two breaking modes such that increasing a breaking
threshold decreases the contribution of the corresponding breaking mode. 
In the beam dynamics shear and bending are coupled so that the bending
breaking
 mode in Eq.\ (\ref{eq:breaking}) mainly captures the shearing of the particle
contacts.
In the model there is only structural disorder present, i.e.\ the breaking thresholds 
 $\varepsilon_{th}$ and $\Theta_{th}$ are set to the same constant values for
all the beams: 
$\varepsilon_{th}=0.003$ and $\Theta_{th}=2^{\rm o}$.
 The breaking criterion Eq.\ (\ref{eq:breaking}) is evaluated in each iteration
step for elongated beams and those ones which fulfill the condition are removed 
from the simulations.
As a result of subsequent beam breaking cracks form in the sample. 

 Those particles which are not connected by beams, either because they have never been
 connected or their beam has broken, interact via Hertz contacts as described in section IIA
 for the packing generation \cite{poschel_grandyn_2005}. 
At this stage of modelling, particles themselves do not fragment, 
and only the connecting beams can be broken. For the material parameters of beams and particles
such as Young modulus, Poisson ratio, and damping constants the same values are used as 
in Refs.\ \cite{carmona_fragmentation_2008,PhysRevE.86.016113}. 

\section{Strain controlled uniaxial loading}
\begin{figure}
\begin{center}
\epsfig{ bbllx=65,bblly=10,bburx=690,bbury=500,file=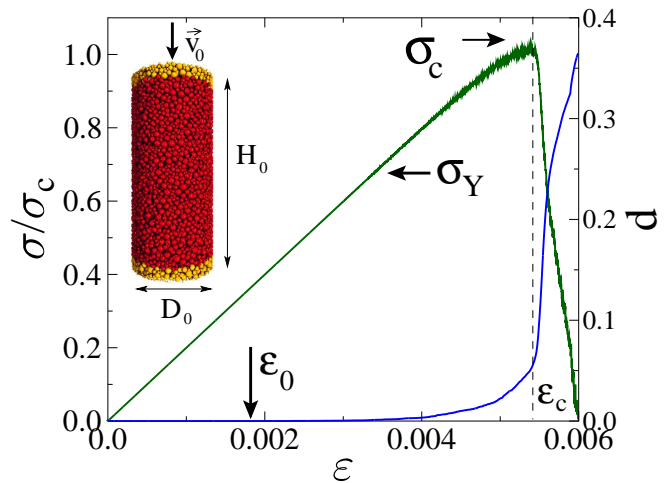,
width=8.5cm}
\end{center}
\caption{
(Color online) Constitutive behavior $\sigma(\varepsilon)$ of a single 
sample (left axis) together with the accumulated damage $d(\varepsilon)$ (right axis). 
The large arrow
and the vertical dashed line indicate the onset of beam breaking $\varepsilon_0$
and the position of the maximum stress $\sigma_c$ of the constitutive curve at $\varepsilon_c$, 
respectively. The constitutive curve is presented in a dimensionless form by
rescaling the stress with its critical value $\sigma_c$.
The inset illustrates the loading condition such that the bottom and 
top layers, highlighted by gold (light) color, are clamped. The top layer moves
downward with velocity $v_0$ while the bottom is fixed.
The visual yield point (discernible departure from linearity) is marked as $\sigma_Y$.
}
\label{fig:constit}
\end{figure}

We carried out computer simulations
of the breaking process of cylindrical specimens of sedimentary rocks 
under uniaxial loading conditions.
Strain controlled loading was performed in such a way that the top two particle
layers on the top and bottom of the sample (a thickness of $2R_{max}$ each) were
clamped rigidly to each other.
Such clamping simulates strong coupling between the platen attached to the
loading piston
in a real experiment and the rock sample. Alternatively a low friction 'shim' is
sometimes
used between the two as an alternate protocol to allow lateral movement of the 
upper and lower layer of the sample. The clamping has the effect of promoting
a shear band to develop between two otherwise largely intact fragments, 
rather than producing a vertical tensile crack or a pervasively shattered
sample.
The bottom layer was fixed while the one on the top was moving downward at a
constant speed $v_0$ which yields a constant strain rate $\dot{\varepsilon}$. 
The value of $\dot{\varepsilon}$ was set as 
$\dot{\varepsilon}\Delta t = 1.8\cdot 10^{-7}$,
where $\Delta t$ is the time step used to integrate the equation of motion.
This is a very common laboratory experimental protocol,
where the constant velocity is known as the piston stroke rate.
The loading condition is illustrated by the inset of Fig.\ \ref{fig:constit}. 
The side wall of the cylinder is completely free, no confining 
pressure was applied. 

The macroscopic axial strain $\varepsilon$ of the sample was obtained as
\beq{
\varepsilon=\Delta H/H_0,
}
where $\Delta H$ is the displacement of the top layer
(see inset of Fig.\ \ref{fig:constit}). In order to characterize the mechanical 
response of the sample
we measure the force $\vec{F}^t$ acting on the top layer, which is needed to
maintain
the deformation $\varepsilon$. The axial stress $\sigma$ is determined as 
\beq{
\sigma = F^t_y/(\pi D_0^2/4),
}
where $F^t_y$ denotes the vertical component of the total force $\vec{F}^t$.
The constitutive curve $\sigma(\varepsilon)$ of the system is presented 
in Fig.\ \ref{fig:constit} together with the cumulative damage $d(\varepsilon)$
defined as the fraction of broken beams.
\begin{figure}
\begin{center}
\epsfig{ bbllx=0,bblly=0,bburx=290,bbury=355,file=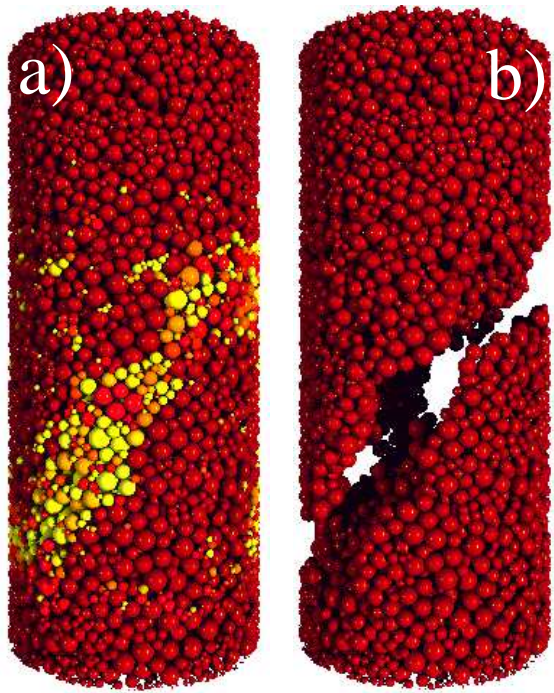,
width=7.5cm}
\end{center}
\caption{
(Color online) $(a)$ Re-assembled sample in the final state of the simulation.
Particles are placed back to their original location in the cylinder and they
are 
colored according to the size of the fragment they originally belonged to.
Hence, the two big residues are dark red, 
single particles have yellow color, while other
colors 
indicate fragment sizes between the two limits. The damage band can clearly be 
observed on the surface of the cylindrical sample.
$(b)$ To have a better view on the damage band, we only present the two big 
pieces of the specimen without the small sized fragments. The width and the
orientation
of the damage band can be better inferred from this representation. 
}
\label{fig:broken_sample}
\end{figure}
The system has a highly brittle response:
for small deformations linearly elastic behavior is obtained,
stronger non-linearity of $\sigma(\varepsilon)$ is only observed in the vicinity
of its maximum. The position $\varepsilon_c$ and value $\sigma_c$ of the maximum stress
characterize the ultimate strength of the sample. 
Deviations from linear elasticity in the stress-strain curve are associated 
with the onset of damage due to beam breaking, indicated by the large vertical arrow 
on Fig.\ \ref{fig:constit}. 
However, there is a large delay before the non-linearity becomes obvious. 
This means that the ‘yield point’ commonly 
identified by the onset of discernible non-linearity 
in the stress-strain curve ($\sigma_Y$ in Fig.\ \ref{fig:constit}), is likely 
to be an overestimate, and that linearity cannot 
be taken as diagnostic evidence of elastic behaviour. 
The fluctuations in stress also become more obvious and grow after $\sigma_Y$, as 
systematically-larger bursts occur in the approach to peak stress. 
For a real laboratory test with a finite detection threshold 
for acoustic emission magnitude set by the ambient noise level, 
this early onset of damage would not be discernible. 
Similarly the stress strain curves are usually 
much smoother in real laboratory tests, even beyond 
the peak stress (e.g.\ similar curves in \cite{sammonds_role_1992,ojala_2004,Heap201171,Mair200025}). 
This could be due, in part at least, to other forms of silent damage not modelled here, 
e.g.\ by environmentally-assisted, quasi-static stress 
corrosion \cite{sabine_PhysRevE.88.032802,danku_PhysRevLett.111.084302}.
The damage continues to increase well
into the period of dynamic stress drop. The simulation stops when the 
force acting on the top layer drops down to zero. The average CPU time needed 
to complete the compression simulation 
of a single sample of 20000 particles is about 5 hours on a single core of an Intel Xeon
(6 cores) processor. 

\section{Formation of a damage band}
During the loading process the weakest contacts break first. Due to the 
structural disorder of the sample, these breaking events give rise to 
micro-cracks randomly scattered all over the sample. Simulations revealed that 
as loading proceeds, 
damage localizes to a narrow spatial region where gradually a high fraction
of beams break and a macroscopic crack emerges spanning the entire cross section
of the sample. Figure \ref{fig:broken_sample} presents a representative example 
of the final breaking scenario of the sample. To have a clear 
view on the localized damage, the sample is reassembled in such a way 
that the particles are placed back to their original position and they are 
colored according to the size of the fragment they belong to. 
By fragments here we mean both individual particles whose beams are all broken, 
or clusters of two or more particles still connected by at least one beam.
Figure \ref{fig:broken_sample}$(a)$ shows the complete cylindrical sample, while
in Fig.\ 
\ref{fig:broken_sample}$(b)$ only the two big intact fragments or 'residues' are
highlighted. 
One can clearly observe the geometrical shape of the damage band which has a 
well defined orientation. The damage band comprises
a large number of small sized fragments, i.e.\ individual particles
and small fragments composed of a few particles. 
The quantitative analysis of $740$ samples from the different model runs
revealed that the angle 
of the deformation band with respect to the load direction falls always between 
$30$ and $45$ degrees. This is similar to the angle measured in real
experiments,
implying a coefficient of internal friction between 0 and 0.7. In nature
real faults are typically aligned with an angle of 30 degrees or so to the 
maximum principal stress \cite{paterson_book_1978}, although much lower 
coefficients of internal friction
can be inferred in materials such as unconsolidated water-saturated clay.  

The damage band gradually evolves and its final width reaches some
$5-8$ average particle diameters. The high concentration of damage
implies that the majority of beams are broken inside the band, however,
not all of them are. Inside the deformation band 
spherical particles connected by the surviving intact beams form fragments which 
are embedded into the background of single particles (fine powder in the model). 
These fragments emerge 
as the width of the band increases by gradually crushing the major pieces
along the sheared faces. We determined the mass distribution $p(m)$ of 
fragments averaging over 600 simulations. In Figure \ref{fig:massdist}
three regimes of the fragment mass distribution can be distinguished:
\begin{figure}
\begin{center}
\epsfig{ bbllx=30,bblly=20,bburx=375,bbury=330,file=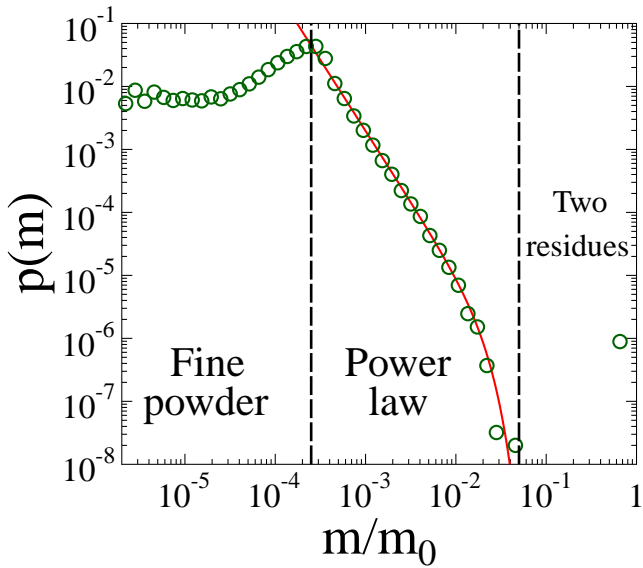,
width=8.5cm}
\end{center}
\caption{
(Color online) Mass distribution of fragments in the final state of the 
simulations. Three regimes can be distinguished: the two big residues
illustrated in Fig.\ \ref{fig:broken_sample}$(b)$
give rise to the single peak at about a few tenths of the total sample mass
$m_0$. 
A large fraction
of fragments forms the fine powder which comprises single particles
and very small fragments of a few particles. In the medium range
the fragments have a power law distribution with a stretched exponential
cutoff. 
The red line represents the fit obtained with Eq.\ (\ref{eq:fitpower}).
}
\label{fig:massdist}
\end{figure}
The two big residues give rise to the peak at about a few tens of the 
total mass of the sample $m_0$. The majority of fragments form 
the fine powder on the left of the figure. These fragments comprise
a single particle or a few particles up to 3-4, hence, their mass
distribution is mainly determined by the original size distribution of
the sedimented spherical particles.
The most remarkable result is that in the intermediate range 
the fragments have a power law mass distribution over nearly
two orders of magnitude. The numerical results were fitted by the functional
form 
\beq{
p(m) \sim m^{-\tau}\exp{\left[-(m/m^*)^{\mu}\right]},
\label{eq:fitpower}
}
where the cutoff has a stretched exponential form.
A best fit in Fig.\ \ref{fig:massdist} was obtained with the parameter 
values $\tau=2.1$, $\mu=2.0$, and $m^*/m_0=0.015$. A power law distribution of 
fragment sizes is also typically observed in fault wear products (gouge) in natural 
and laboratory faults 
\cite{sammis_pag_1987,Biegel1989827,turcotte_factals_1986,steacy_automaton_1991,
turcotte_fractals_1997}.
Several mechanism-based theoretical models have been 
proposed where the power law size distribution is the direct consequence of the
scale-invariance built into the breaking mechanism
\cite{steacy_automaton_1991,turcotte_fractals_1997}.
In our DEM the gradual compression of the specimen naturally leads to the
emergence
of a power law distribution in the intermediate mass range. This may imply that
after the 
localization of damage the subsequent broadening of the deformation band by
gradually
crushing the surrounding material involves some degree of self similarity. 
This broadening of the shear band and relatively uniform geometric properties 
of its contents has also been observed in laboratory experiments on natural 
sandstones \cite{Mair200025} and has been validated independently by modeling 
and measurement of the evolution of fluid permeability across the band 
\cite{main_fault_2000}.
In real laboratory and experimental faults of the frequency-length distribution
for particles of fault gouge measured by a laser particle size analyser is 
$N(l)\sim l^{-D}$, with an exponent $D$ around $2.6$ \cite{turcotte_fractals_1997,
sammis_pag_1987,Biegel1989827,turcotte_factals_1986,steacy_automaton_1991}. The
probability-density distribution exponent for length is then $3.6$. For spherical
particles of constant density, as here, the volume $V\sim m\sim l^3$ so
$p(m)=p(l)dl/dm\sim m^{-(1+D)/3}m^{1/3-1}\sim m^{-\tau}$, where $\tau=1+D/3$.  If $D=2.6$ 
then $\tau=1.87$, compared to $2.1$ in our simulations.  The remaining discrepancy 
is likely due to the angularity (non-sphericity) of real fault gouge \cite{turcotte_fractals_1997}, 
when mass will not necessarily scale with the cube of linear dimension, to differences 
in density for the different minerals involved, and perhaps to the fact that individual 
particles cannot be crushed in the model.

 \section{Crackling noise during fracture}
 In spite of the relatively smooth macroscopic response of the system presented in Fig.\
\ref{fig:constit}, 
 at the micro-level the breaking of beams evolves in an intermittent way. 
 The structural heterogeneity of the material has an important effect
 on the beginning of the breaking process which is dominated by the uncorrelated
 nucleation of micro-cracks. As the strain controlled loading proceeds
consecutive
 beam breakings become correlated: the stress released by broken beams gets
 redistributed in the surroundings which can induce additional breakings 
 and in turn can trigger an entire avalanche of local breaking events.
 These correlated trails of local beam breakings or ‘bursts’ can be 
 considered analogous to acoustic
 emissions generated by the nucleating and propagating cracks in laboratory
experiments 
 \cite{lockner_1993,carpinteri_criticality_2009,sammonds_role_1992}.
 
 In our DEM simulations the breaking criterion Eq.\ (\ref{eq:breaking})
 is evaluated in each iteration step of the equation of motion of the system
 and those beams which fulfill the condition get broken. As loading proceeds, 
we record the breaking time 
 $t_i^b$ of each beam $i=1,\ldots , N_b$, where $N_b$ is the total number of
 broken beams. In order to quantify the temporal correlation of local breaking
 events and the jerky cracking process induced by the subsequent load
redistribution, 
 we introduce a correlation time $t_c$: if two consecutive beam breakings occur 
 within the correlation time $|t_{i+1}^b-t_i^b|<t_c$ they are considered to
belong 
 to the same burst. The value of $t_c$ was set to $t_c=25\Delta t$, which is 
 approximately the time needed for the elastic waves to pass the radius of the 
 sample $D_0/2$. 
 The concept of bursts defined as correlated trails of elementary events 
 has also been used recently to study intermittent
 human activity based on telecommunication data \cite{kaski_kertesz_2012}. 

The size of a burst $\Delta$ is defined as the number of beams breaking in
the correlated sequence. Since beams represent particle contacts, the
size $\Delta$ is related to the total rupture area created by the burst. 
Based on the above definition further useful quantities can be introduced 
to characterize the bursting activity accompanying the breaking process:
The time of occurrence $t$ of a burst of size $\Delta$ is the 
arithmetic average of the time of the first and last beam breakings 
in the burst
\beq{
t = (t_{\Delta}^b + t_1^b)/2.
} 
The burst duration $T$ is the difference of the time of the first and last beam 
breaking in the burst 
\beq{
T = t^b_{\Delta} - t^b_{1}.
}
The elastic energy $E_j^b$ stored in the deformation of beams
gets released during the breaking event. The total energy $E$ 
released by a burst can be determined as the sum of beam energies
\beq{
E = \sum_{j=1}^{\Delta_i}E_j^b.
} 
The waiting time $t_w$ between two consecutive bursts of size $\Delta_i$ and 
$\Delta_{i+1}$ is the difference of the time of the first beam breaking of the 
second burst and the time of the last breaking event of the first one 
\beq{
t_w = t_1^{b(\Delta_{i+1})} - t^{b(\Delta_i)}_{\Delta_i}.
}
We have seen in the previous section that the damage has an emergent, 
highly non-trivial spatial structure 
in the sample. Consequently, the spatial positioning of bursts provides
also valuable information on the evolution of the fracture process. 
The spatial position $\vec{r}$ of a burst can be 
characterized by calculating the center of mass position of broken beams
\beq{
\vec{r} = \frac{\sum_{j=1}^{\Delta} \vec{r}_j^b}{\Delta},
}
where $\vec{r}_j^b$ is the position of beam $j$ broken in 
the avalanche.

\begin{figure}
\begin{center}
\epsfig{ bbllx=30,bblly=10,bburx=410,bbury=305,file=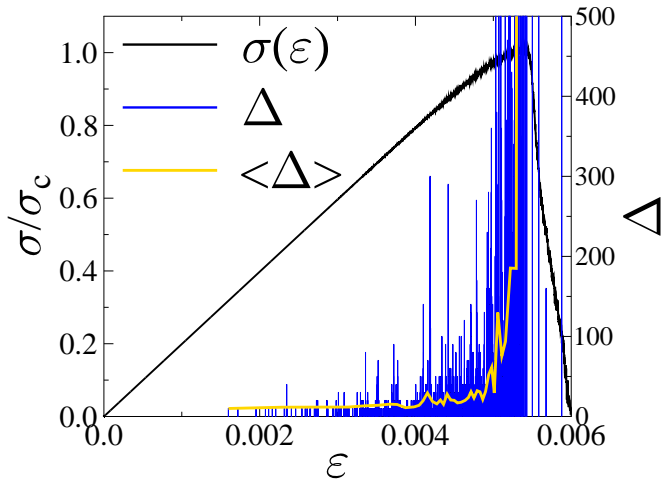,
width=8.5cm}
\end{center}
\caption{
(Color online) Time series of burst size $\Delta$ obtained from a single model run
as a function of the strain of their appearance 
together with the corresponding constitutive curve.
The burst size $\Delta$ has strong fluctuations and on the average it
increases when the maximum of $\sigma(\varepsilon)$ is approached.
The moving average of $\Delta$ was calculated over 50 consecutive events.
}
\label{fig:timeseries}
\end{figure}

Figure \ref{fig:timeseries} demonstrates the size of bursts $\Delta$
as a function of strain $\varepsilon$ of their appearance together with its
moving
average. For comparison
the constitutive curve $\sigma(\varepsilon)$ of the system is also presented.
In spite of the smooth curve of the accumulated damage 
$d(\varepsilon)$ in Fig.\ \ref{fig:constit} the size of bursts $\Delta$ exhibits
strong fluctuations while its average increases as the maximum of the 
constitutive curve $\sigma(\varepsilon)$ is approached. 
At the beginning of the breaking process only small bursts of a few breaking 
beams appear, however, as loading proceeds the triggering of longer avalanches
becomes
more probable. Strong bursting activity with complex structure of the event
series
emerges after the value of $\sigma$ exceeds approximately the two third of
the peak stress $\sigma_c$. 
The maximum of the axis of $\Delta$ in Fig.\ \ref{fig:timeseries}
is set to 500 to be able to see the details of small sized events as well.
The maximum burst size in the example is $\Delta_{max}=23417$ which was reached
slightly after the peak of $\sigma(\varepsilon)$ followed by a few additional 
large avalanches with size $\sim 1000$ connected beam breakings.
Such large sized events with long duration are the consequence of the 
formation of the damage band where long breaking sequences emerge due to
intense fragmentation of the sample in the damage band. 
With the present value of the correlation time 
$t_c$ the total number of bursts we identify during the fracture process
of a single sample is about 2000. 

Recently, we have carried out a detailed analysis of the statistics of
the size, energy, and duration of bursts, furthermore, of the waiting
times between consecutive events \cite{ourpaperinprl}. 
In this reference 
we considered averages integrated over the whole test before and after 
catastrophic failure, respectively.
We showed that all quantities are power law distributed
with an exponential cutoff. Careful testing of the burst identification revealed 
that varying the value of the 
correlation time $t_c$ only affects the cutoff. 
The functional form and the exponents of the distributions
remained robust until $t_c$ falls close to the time needed for the elastic wave to
pass the specimen.
Single burst quantities proved to be correlated,
i.e.\
bursts of larger size typically release a higher amount of energy and have a
longer 
duration. These correlations are quantified by the power law dependence of the
average energy 
and duration on the burst size \cite{ourpaperinprl}. 
In the following we focus on the space-time variation
of the crackling activity in order to understand how the slowly driven system
approaches
macroscopic failure.

\section{Approach to failure}
The beginning of the fracture process is dominated by the structural disorder
of the material resulting in random nucleation of small cracks comprising
only a few broken beams all over the sample. However, the subsequent stress
redistribution around cracks gives rise to correlations between failure events
which becomes more and more relevant as the system approaches the maximum 
of the constitutive curve. 

\subsection{Event size statistics}
\begin{figure}
\begin{center}
\epsfig{
bbllx=10,bblly=10,bburx=380,bbury=610,file=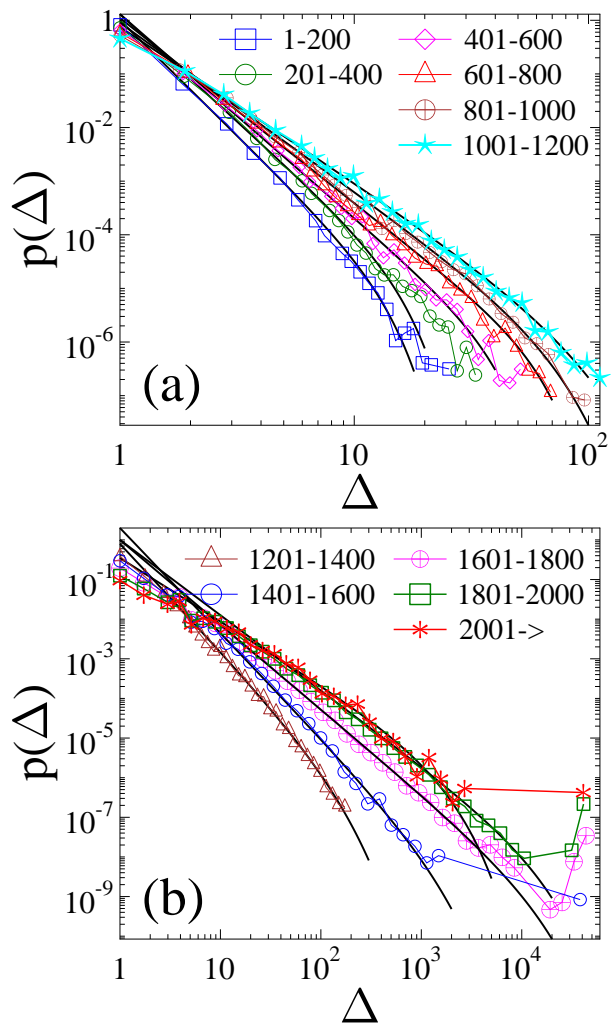,
width=8.5cm}
\end{center}
\caption{
(Color online) Size distribution of bursts $p(\Delta)$ calculated in windows of 
200 consecutive events in the time series without overlap. The thick continuous lines
represent fits with Eq.\ (\ref{eq:b-value-form}). 
}
\label{fig:b-value-200}
\end{figure}
\begin{figure}
\begin{center}
\epsfig{ bbllx=40,bblly=40,bburx=420,bbury=330,file=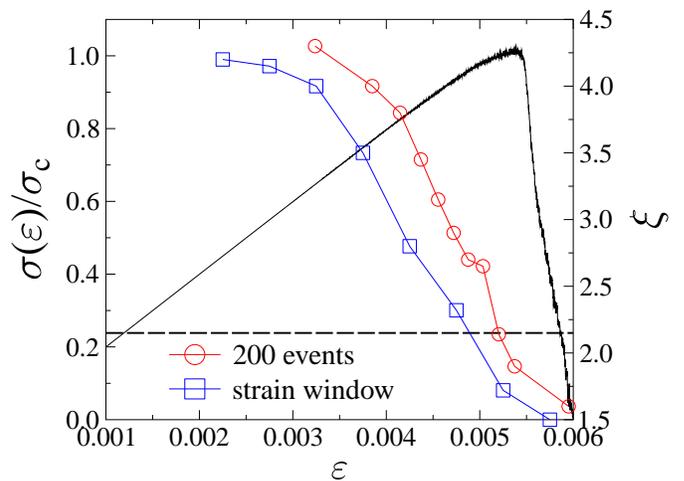,
width=8.5cm}
\end{center}
\caption{
(Color online) The exponent $\xi$ of the burst size distribution evaluated for 
windows of 200 consecutive events and for equidistant bins of strain during the 
loading process. The dashed horizontal line represents the value of $\xi$
obtained
by taking into account all the events. For clarity, we also show a
representative 
example of the constitutive curve.
}
\label{fig:exponents}
\end{figure}

Accumulating all the events up to failure, we have shown \cite{ourpaperinprl}
that the 
size distribution $p(\Delta)$ of bursts can be well described by 
the functional form
\beq{
p(\Delta) \sim \Delta^{-\xi}\exp{\left[-\left(\Delta/\Delta^*\right)^c\right]}.
\label{eq:b-value-form}
} 
The value of the exponent $\xi$ was obtained numerically as $\xi=2.22$.
In order to investigate if the exponent $\xi$ of the size distribution
$p(\Delta)$ 
of bursts depends on when bursts were generated during the loading process
the following analysis was carried out: we evaluated the size distribution
$p(\Delta)$
in windows of 200 consecutive events without any overlap. Since the total number
of
bursts falls between 1800 and 2200 for each sample, 11 event windows could be 
analyzed which were averaged over 600 samples.  
In Figure \ref{fig:b-value-200} the results for the first 6 and the last 5
windows 
are presented separately to have a more clear view of the details. 
In the consecutive windows the average size of bursts increases, however,
the functional form of the distribution remains nearly the same as that given 
by Eq.\ (\ref{eq:b-value-form}).
Careful analysis of the data showed that rescaling the distributions by some
powers 
of the corresponding average burst size along the horizontal and vertical 
axis no data collapse can be achieved. The result implies that the change of
$p(\Delta)$ 
in Fig.\ \ref{fig:b-value-200} cannot be explained by the changing cutoff, but
the exponent $\xi$ depends on the position of the event window in the time
series.
In Fig.\ \ref{fig:exponents} we present the exponent $\xi$ obtained by maximum
likelihood
fitting with Eq.\ (\ref{eq:b-value-form}) as a function of the average value of
the strain 
where bursts occurred in a given event window.
The most remarkable feature of the results is that the value of the exponent 
$\xi$ spans a broad
interval monotonically decreasing from $\xi=4.3$ down to $\xi=1.6$
for events emerging beyond the peak of the constitutive curve. 
The smaller value of $\xi$ indicates that approaching the maximum of 
$\sigma(\varepsilon)$ the relative frequency of large events increases
in the windows. This is clearly visible in Fig.\ \ref{fig:b-value-200}.

The average strain of events can be slightly misleading especially 
for windows at the beginning of the breaking process, since 
here the strain of events may span a broad range with an uneven 
distribution. In order to avoid this problem, we also analyzed 
the size distribution of events in equally sized strain bins 
$\Delta \varepsilon=0.0005$.
The functional form of the distributions on Fig.\ \ref{fig:b-value-strain}$(a)$
is the same as before and can be well described with a power law followed 
by an exponential cutoff as in Eq.\ (\ref{eq:b-value-form}).
The value of the exponent $\xi$ obtained by fitting with 
Eq.\ (\ref{eq:b-value-form}) is also plotted in Fig.\ \ref{fig:exponents}
at the middle points of the strain windows. 
In both cases the exponent spans practically the 
same range and have a similar dependence on the position of the measurement 
along the loading process. In Fig.\ \ref{fig:exponents} 
the horizontal dashed line indicates the value of the average burst size exponent 
$\xi=2.22$ obtained when all events are considered in the statistics. 
Comparing the three exponents one can conclude that the asymptotics 
of the distribution when all events are considered is dominated by the strongly
non-linear 
regime of the constitutive curve before reaching the maximum.

Before comparing these results with those of laboratory tests it is necessary to discuss the
way the acoustic emissions are measured, and how they scale with energy and source size. Typically
the frequency-size distribution for natural earthquakes and acoustic emissions is characterised by
the Gutenberg-Richter law for small and intermediate sized events: $\log(N)=a-bm$, where $a$ and $b$ are
empirical constants, $N$ is an incremental frequency, and the magnitude $m$ is determined from the common
logarithm of the peak amplitude of the radiated wave, corrected for attenuation with distance to the
source location, so that $1$ magnitude unit is equivalent to $20dB$ in acoustics. From a simple
dislocation model for the source, the relationship between magnitude and energy is $\log E = c + dm$,
where $d=3/2$ if energy scales with source dimension (area) $\Delta$ as $\sim \Delta^{\nu_E}$, 
with $\nu_E=3/2$ (see e.g.\ Ref.\ \cite{turcotte_fractals_1997}). We have previously 
shown $\nu_E = 1.15$ (Ref.\ \cite{ourpaperinprl}) in our simulations rather than $3/2$. The difference
is likely because the cascades of broken bonds are not necessarily planar objects, as assumed in the
simple dislocation theory. From Eq.\ (\ref{eq:b-value-form}) the equivalent relation 
for the power law part is recovered in the form of the Gutenberg-Richter law if $\xi=b+1$. 
From this the typical value of $b$ in Figure \ref{fig:exponents} is $1.22$, 
ranging from $3$ or so at the start of the experiment down to $0.5$ at or near
catastrophic failure. The average is very close to the typical value $b=1$ for natural earthquakes 
\cite{turcotte_fractals_1997},
and for the average in a typical laboratory test, such as the results of Ref.\ \cite{sammonds_role_1992}, 
where the ‘b-value’ ranges from around $1.7$ down to $0.7$. The laboratory tests have 
a more restricted range because they
typically record only the largest events, and many smaller events which would otherwise contribute
to the high value of $b$ early in the test are missing. Quantitatively our results then quite closely
reproduce the monotonic decrease in $b$ observed in the ‘dry’ test on the porous sandstone sample in
Ref.\ \cite{sammonds_role_1992} (their Figure $1(a)$), previously modelled in a more conceptual 
way by simplified cellular automaton models restricted to two dimensions 
\cite{main_fault_2000,Henderson1992905}. At this stage of modelling there is no
coupling with a pore fluid phase, so it is not yet possible to reproduce the more complex fluctuations
in $b$ associated with changes in effective stress as the pore pressure is introduced and varied 
in Ref.\ \cite{sammonds_role_1992}.

\subsection{Spatial statistics}
\begin{figure}
\begin{center}
\epsfig{ bbllx=10,bblly=0,bburx=390,bbury=230,
file=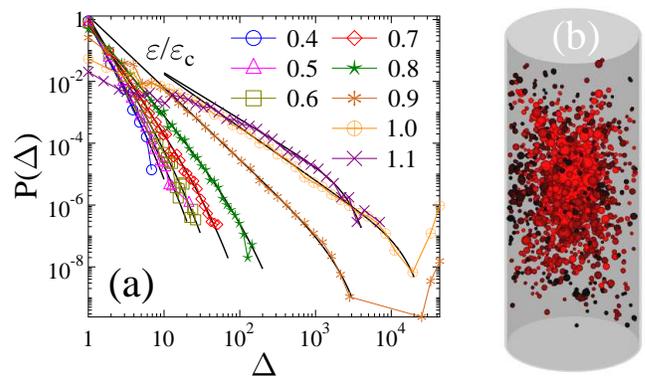,
width=8.5cm}
\end{center}
\caption{
(Color online) $(a)$ The size distribution of bursts evaluated in equally
sized strain bins $\Delta \varepsilon=0.0005$. The functional form of the curves can again be described
by Eq.\ (\ref{eq:b-value-form}). The legend indicates the middle point
of the strain bin divided by the critical strain $\varepsilon_c$. 
$(b)$ Spatial position of bursts in a single
sample. Single bursts
are represented by spheres with a diameter proportional to the size of the
event.
For clarity, only events with size $\Delta > 5$ are shown. The color code
illustrates
the time sequence of events such that dark and light colors stand for early and 
late bursts, respectively.
}
\label{fig:b-value-strain}
\end{figure}

\begin{figure}
\begin{center}
\epsfig{ bbllx=5,bblly=5,bburx=420,bbury=460,file=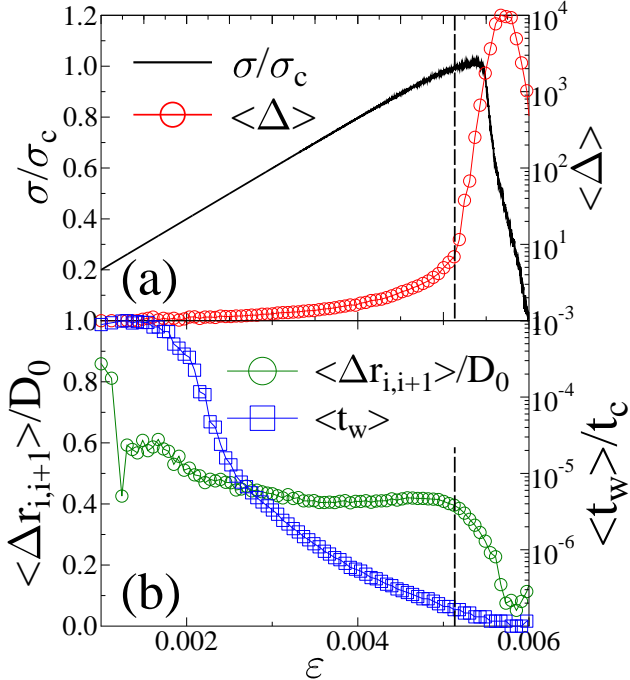,
width=8.5cm}
\end{center}
\caption{
(Color online) $(a)$ average size of bursts $\left<\Delta\right>$ as a 
function of strain $\varepsilon$ together with the constitutive curve 
$\sigma(\varepsilon)$. $(b)$ Average waiting time between consecutive
events $\left<t_w\right>$ and the average distance between 
bursts normalized by the diameter of the base circle of the cylinder as 
function of strain. 
}
\label{fig:position}
\end{figure}

By locating acoustic emissions and slowing down the failure process 
\cite{lockner_nature_1991} demonstrated that damage localization typically
occurs near (sometimes a little before) the peak stress. 
Here a similar behavior is seen in the simulation results.
In Fig. \ref{fig:b-value-strain}$(b)$ the spatial position of bursts with size 
$\Delta >5$ are presented in a single sample.
Small events are scattered all over the sample, however,
the large ones which occur in the vicinity of the peak load $\sigma_c$ 
are localized to a limited volume.
In order to quantify how this localization develops, we calculated the
average distance $\left<\Delta r_{i,i+1} \right>$ of consecutive bursts as a
function of strain $\varepsilon$
\beq{
\left<\Delta r_{i,i+1} \right>=\sqrt{\left<(\vec{r}_{i+1}-\vec{r}_i)^2\right>},
}
where $\vec{r}_i$ and $\vec{r}_{i+1}$ are the center of mass positions 
of two consecutive bursts. In Figure \ref{fig:position}$(b)$ the value
of $\left<\Delta r_{i,i+1} \right>$ is normalized by the initial diameter $D_0$
of the 
cylindrical sample. At the beginning 
of the fracture process the average distance is practically equal
to the half of the sample diameter which is the consequence of the absence
of correlations. In this regime bursts are randomly scattered without any
apparent correlation. The vertical dashed line marks the strain value
where the average distance between burst locations starts to decrease rapidly - a clear signature of
the emerging correlation.
For comparison in Fig.\ \ref{fig:position}$(a)$
we also present the constitutive curve together with the average size of 
bursts $\left<\Delta\right>$. The figure shows that in the uncorrelated 
regime bursts are 
relatively small $\Delta < 10$, while the onset of the correlated burst 
regime is also accompanied by the sudden increase of the average burst size. 
The average waiting time $\left<t_w \right>$ monotonically decreases in Fig.\ 
\ref{fig:position}$(b)$ with increasing strain showing the acceleration of cracking.

\begin{figure}
\begin{center}
\epsfig{ bbllx=10,bblly=10,bburx=365,bbury=315,file=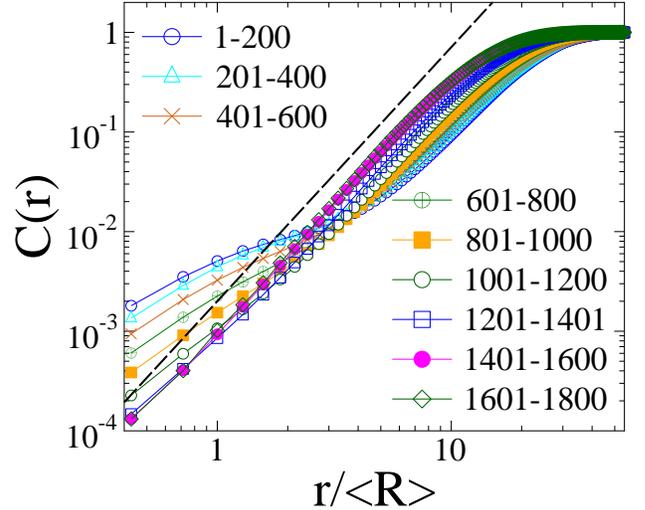,
width=8.5cm}
\end{center}
\caption{
(Color online) Correlation integral $C(r)$ for windows of $n=200$ consecutive
events.
The straight line represents a power law of exponent $2.55$. 
}
\label{fig:correl}
\end{figure}

To obtain a more detailed description on how the spatial correlation 
of bursts evolves we determined the 
correlation integral $C(r)$ for windows of $n=200$ consecutive events similar to
the b-value analysis of the size distribution of events in the previous section. 
The value of $C(r)$ was determined by counting the number of pairs of bursts
$N(<r)$ which are separated by a distance less than $r$
\beq{
C(r) = \frac{N(<r)}{N_p},
} 
which was normalized by the total number of pairs $N_p=n(n-1)/2$ of the $n$
events of the window.
In Figure \ref{fig:correl} the distance of bursts have a lower and upper cutoff
which are
determined by the size of the particles and by the size of the sample,
respectively.
At high distances $C(r)$ converges to $1$ due to normalization.
At the beginning of the fracture process $C(r)$ only 
saturates when $r$ spans the sample $r\approx 50\left<R\right>$ 
since the $n$ bursts are scattered over the complete volume. 
However, in the vicinity of the peak load $\sigma_c$, i.e.\ for the last 4
event 
windows, a power law behavior of $C(r)$ emerges
\beq{
C(r) \sim r^{D_2},
}
where the correlation dimension is $D_2=2.55$.

This value of the correlation dimension, and the fact that the slope decreases systematically
as failure approaches, compares well with the results obtained from laboratory tests on Oshima
Granite \cite{GJI:GJI369} who found $D_2$ to be $2.75$, $2.66$ and $2.25$
for the primary, secondary and tertiary creep phases in a constant load (creep) test.

\section{Conclusions}
We presented a discrete element investigation of
the fracture of sedimentary rocks under uniaxial compression focusing on how
the 
system approaches macroscopic failure. 
The problem has a high relevance to understand the
emergence of catastrophic failure in cohesive granular materials, including the failure of building
materials and natural or induced events such as landslides or earthquakes occurring in sedimentary
rocks, where compressive failure plays a crucial role.
To generate the heterogeneous micro-structure of porous rocks, spherical
particles
are sedimented in a cylindrical container with a log-normal size distribution.
The repulsive interaction of particles is captured by Hertz contacts, while
cohesion is provided by beam elements representing the cementitious coupling of
particles.
Breaking of beams is induced by stretching and shearing combined in a
physical breaking rule. Strain controlled uniaxial loading is realized 
by clamping the two opposite ends of the sample one of which was slowly moved
at a constant velocity. 

Since clamping of the sample ends promotes shear failure, damage strongly 
concentrated in a band. Inside the damaged band the material is mainly 
heavily fragmented into a fine powder of single particles or clusters of 
at most a few still-connected particles, however it also contains larger
fragments
with a power law mass distribution. The results are in a good qualitative
agreement with 
the size distribution of fault wear products both in natural and laboratory
faults
\cite{turcotte_factals_1986,turcotte_fractals_1997,steacy_automaton_1991}. 
The power law behavior imply that the gradual cracking of the sample as the
damage 
band broadens involves some degree of self-similarity. The value of the
exponent 
$\tau=2.1$ is higher than the one found in DEM simulations of the impact induced
dynamic 
breakup of heterogeneous materials in three dimensions $\tau=1.8-1.9$ 
\cite{carmona_fragmentation_2008,timar_new_2010,PhysRevE.86.016113}. It
indicates
that slow crushing gives rise to a lower frequency of large fragments since 
the major fraction of the body is comprised by two nearly intact residues.
 
Simulations revealed that the process of damage accumulation is not smooth,
instead it is composed of cascades of microscopic breakings triggered 
by the subsequent load redistribution after local failure events. Such
intermittent bursts
of breakings are directly analogous to sources of acoustic emissions in real
experiments.
Here we focused on how the statistics of burst sizes and the correlation of the
spatial 
location of bursts evolve as the system approaches macroscopic failure. 
Considering non-overlapping windows of 200 consecutive events we showed that the
size
of bursts is power law distributed with an exponent $\xi$ which decreases to
$\xi \approx 1.6$
towards failure in a way that is quantitatively very similar to
that seen in laboratory tests on sandstone samples. This so-called {\it b-value anomaly} 
has been observed both in
laboratory
experiments \cite{sammonds_role_1992,hatton_1993,ojala_2003} and in field
measurements on earthquakes, 
however, most of
the simpler modeling approaches applied to date have failed to reproduce this behavior
of the avalanche statistics. For the fiber bundle model
it was pointed out analytically in the mean field limit that gradually
restricting 
the data evaluation to the close proximity of the failure point the size
distribution of
breaking avalanches exhibits a crossover from the exponent 5/2 to a lower value
3/2 \cite{pradhan_crossover_2005-1}.
Later on computer simulations in the limit of localized load sharing yielded the
same 
crossover behavior \cite{raischel_local_2006}. However, 
the nature of this crossover is different from the one responsible for the 
b-value anomaly: in FBMs the exponent of the avalanche size distribution is
always 3/2
whenever events are considered in a narrow enough window irrespective of the
stress level.
An apparent crossover of $p(\Delta)$ was reported in a lattice model of the
compressive 
failure of heterogeneous materials
\cite{amitrano_rupture_2006,amitrano_2012_epjst}.
However, it was shown that rescaling the distributions with powers of the
average burst size
data collapse can be achieved, which clearly shows that the exponent is
constant.

Our simulations revealed that the beginning of the failure process is dominated
by 
the quenched disorder of the material so that small sized bursts randomly pop up
all over the sample. In the vicinity of macroscopic failure spatial correlations
emerge:
the average distance of consecutive bursts sets to a rapid decrease and the
correlation 
integral of events of non-overlapping windows develops a broad power law regime.

\begin{acknowledgments}
The work is supported by the projects TAMOP-4.2.2.A-11/1/KONV-2012-0036, 
TAMOP-4.2.2/B-10/1-2010-0024, and ERANET\_HU\_09-1-2011-0002. The
project is implemented through the New Hungary Development Plan,
co-financed by the European Union, the European Social Fund and 
the European Regional Development Fund.
F.\ Kun acknowledges the support of OTKA K84157.
This work was supported by the European Commissions by the
Complexity-NET pilot project LOCAT.
\end{acknowledgments}

\bibliography{/home/feri/papers/statphys_fracture}

\end{document}